\documentclass[useAMS,usenatbib,onecolumn]{mn2e}
\usepackage{epsfig,amssymb,amsmath,graphicx,enumitem}

\newcommand{\be}{\begin{equation}}
\newcommand{\ee}{\end{equation}}
\newcommand{\er}{{\bf{\hat{e}}}_r}
\newcommand{\ephi}{{\bf{\hat{e}}}_\phi}

\newcommand{\etht}{{\bf{\hat{e}}}_\theta}

\newcommand{\rstar}{R_*}
\newcommand{\mstar}{M_*}
\newcommand{\bmax}{B_{\mathrm{max}}}
\newcommand{\blmax}{B_{l,\mathrm{max}}}
\newcommand{\msun}{M_\odot}

\title[Neutron star deformation due to arbitrary-order multipolar magnetic fields]{Neutron star deformation due to multipolar magnetic fields}
\author[A. Mastrano, P. D. Lasky, and A. Melatos]{A. Mastrano\thanks{E-mail:
alpham@unimelb.edu.au}, P. D. Lasky\thanks{E-mail: paul.lasky@unimelb.edu.au}, and A.
Melatos\thanks{E-mail: amelatos@unimelb.edu.au}\\School of Physics, University of Melbourne, Parkville VIC
3010, Australia}

\begin{document}

\date{Accepted ?. Received ?; in original form ?}

\pagerange{\pageref{firstpage}--\pageref{lastpage}} \pubyear{?}

\maketitle

\label{firstpage}

\begin{abstract}

\noindent{Certain multi-wavelength observations of neutron stars, such as intermittent radio emissions from rotation-powered pulsars beyond the pair-cascade death line, the pulse profile of the magnetar SGR 1900+14 after its 1998 August 27 giant flare, and X-ray spectral features of PSR J0821$-$4300 and SGR 0418+5729, suggest that the magnetic fields of non-accreting neutron stars are not purely dipolar and may contain higher-order multipoles. Here, we calculate the ellipticity of a non-barotropic neutron star with (i) a quadrupole poloidal-toroidal field, and (ii) a purely poloidal field containing arbitrary multipoles, deriving the relation between the ellipticity and the multipole amplitudes. We present, as a worked example, a purely poloidal field comprising dipole, quadrupole, and octupole components. We show the correlation between field energy and ellipticity for each multipole, that the $l=4$ multipole has the lowest energy, and that $l=5$ has the lowest ellipticity. We show how a mixed multipolar field creates an observationally testable mismatch between the principal axes of inertia (to be inferred from gravitational wave data) and the magnetic inclination angle. Strong quadrupole and octupole components (with amplitudes $\sim 10^2$ times higher than the dipole) in SGR 0418+5729 still yield ellipticity $\sim 10^{-8}$, consistent with current gravitational wave upper limits. The existence of higher multipoles in fast-rotating objects (e.g., newborn magnetars) has interesting implications for the braking law and hence phase tracking during coherent gravitational wave searches.}

\end{abstract}


\section{Introduction}

%The external magnetic field of a neutron star is readily inferred from its spin-down rate, but its internal field is not directly observable. The 1998 August 27 giant flare from the soft gamma-ray repeater (SGR) 1900+14 has been interpreted to imply the existence of a strong internal magnetic field \citep{i01}. The giant flare, which released $\sim\ 10^{37}$ J of energy as X-rays, was accompanied by a 2.3-fold increase in the spin-down rate \citep{metal99,wetal99,tetal00}. \citet{i01} proposed that the flare and the enhanced spin down were caused by a global reconfiguration of the internal magnetic field of $\sim 10^{13}$ T, well above the external dipole field of $6.4\times 10^{10}$ T\footnote{Throughout this paper, we use SI units: 1 T $= 10^{4}$ G.}.

Neutron star magnetic fields are approximately dipolar at (i) radio emission altitudes [leading to S-shaped radio polarization swings \citep{lm88,cm11,bm13} and the pulse-width-period relation \citep{r93}] and (ii) in the outer magnetosphere, where high-energy emission is produced \citep{ry95,lom12}. For millisecond pulsars, \citet{a93} calculated the surface strength of non-dipolar components to be $\lesssim 40\%$ of the dipole. However, some observations, such as the anomalous braking index of some radio pulsars \citep{bt10}, intermittent radio emission from pulsars beyond the pair-cascade `death line' \citep{ymj99,cetal00,gm01,metal03,ml10},\footnote{An alternative explanation for these pulsars beyond the death line involves an `offset' dipole, see \citet{a00}, \citet{bm13}, and references therein. We do not consider the offset dipole model in this paper.} cyclotron resonant scattering line energies of some accretion-powered X-ray pulsars \citep{n05}, the pulse profile of SGR 1900+14 following its 1998 August 27 giant flare \citep{fetal01,td01,tlk02}, and X-ray spectral features of PSR J0821$-$4300 \citep{gha13} and SGR 0418+5729 \citep{gog08,ggo11}, have been taken to indicate the presence of higher-order multipoles close to the surface. Furthermore, while the external magnetic field of a neutron star is readily inferred from its spin-down rate, its internal field is not directly observable and may be composed of high-order multipoles too. Activity in magnetars, e.g., giant flares from soft gamma-ray repeaters (SGRs), has been interpreted to imply the existence of a strong, readjusting internal magnetic field \citep{i01}, and simulations indicate that this internal field may be in a `twisted torus' configuration \citep{bn06,bs06}. In previous papers \citep{metal11,mm12}, we showed how gravitational wave observations constrain the internal field strength. In this paper, we discuss the effects of non-dipolar geometries on the deformation of neutron stars and, hence, their gravitational wave emission.

%\footnote{Throughout this paper, we use SI units: 1 T $= 10^{4}$ G.}

It is well known that a magnetic field deforms a star \citep{cf53,f54,g72,k89,pm04,hetal08,metal11}. Neutron stars, with their intense fields, therefore possess significant ellipticities under certain circumstances, making them good candidates for gravitational wave sources \citep{bg96,mp05,setal05,hetal08,detal09}. Observational upper limits from gravitational waves can be used to set upper limits on stellar ellipticity. Ellipticity, which is roughly proportional to the magnetic energy \citep{c02,hetal08,detal09}, can thus be used to constrain the strength and topology of a star's internal field \citep{c02,detal09,aetal10,metal11,p11}.

%\citet{metal11} constructed hydromagnetic equilibria for stratified, non-barotropic stars. The barotropic assumption, while simplifying calculations, severely restricts the form of the field that can be `fitted' into the star, e.g., \citet{hetal08} found that the field must vanish at the surface, contrary to observations, and \citet{lj09} and \citet{cfg10} found that only configurations dominated by the poloidal component (poloidal energy $\gtrsim 90\%$ of total) are allowed, contrary to numerical simulations of magnetic field evolution [e.g., \citet{bn06}]. Furthermore, \citet{lj12} found that no magnetic field configuration is stable in a barotropic star, contrary to observations and the numerical simulations of \citet{bn06}. On the other hand, \citet{aetal13} have shown that a non-barotropic star can have a stable mixed poloidal-toroidal magnetic field. By abandoning the barotropic assumption, \citet{metal11} were able to construct a simple, self-consistent hydromagnetic equilibrium with an internal field that can be matched to an external dipole. In their configuration, \citet{metal11} are less restricted in the choices of poloidal and toroidal components; they need not be of any particular relative strengths and are independently adjustable. Incidentally, this means that the magnetic field configuration is independent of the equation of state chosen, cf. \citet{lj09} and \citet{cfg10}.

\citet{metal11} constructed hydromagnetic equilibria for stratified, non-barotropic stars. The barotropic assumption restricts the forms of the poloidal and toroidal components that can be `fitted' into the star. \citet{hetal08} found that the field must vanish at the surface, contrary to observations, and \citet{lj09} and \citet{cfg10} found that only configurations dominated by the poloidal component (poloidal energy $\gtrsim 90\%$ of total) are allowed, contrary to numerical simulations of magnetic field evolution [e.g., \citet{bn06}]. By abandoning the barotropic assumption, and assuming stable radial stratification \citep{p92,rg92,r09,aetal13}, \citet{metal11} were able to construct a simple, self-consistent hydromagnetic equilibrium, with an internal field that can be matched to an external dipole (so it can be related directly to observations of the external field), keeping the relative strengths of the poloidal and toroidal components independently adjustable. A purely poloidal \citep{mt73,w73} or a purely toroidal \citep{t73} magnetic field is unstable, but theoretical calculations and numerical simulations \citep{bn06} suggest that a magnetic field with both poloidal and toroidal components is stable over dissipative time-scales (i.e., much longer than the Alfv\'{e}n time-scale). Because a non-barotropic star allows arbitrary poloidal and toroidal field strengths, it can easily accommodate the strong internal fields  which, as suggested by the numerical simulations of \citet{bn06}, are required to stabilise the star.

% [to facilitate comparison with our previous papers, we will use the hydrostatic equilibrium proposed by \citet{metal11}]

In this paper, we show how the ellipticity calculation of \citet{metal11} can be generalised to higher multipoles. An axisymmetric magnetic field of a particular configuration is chosen, the density perturbation induced by this field is calculated, and the ellipticity is calculated from the density perturbation. We show how, in principle, gravitational wave observations constrain the relative strengths of the internal magnetic multipoles. In Sec. 2, we briefly describe how the unmagnetised hydrostatic equilibrium state is chosen and how the density perturbation is calculated. In Sec. 3, we recap briefly the results of \citet{metal11} for a dipole poloidal-plus-toroidal magnetic field and show how the analysis can be extended to add an axisymmetric quadrupole. In Sec. 4, we generalize the work of \citet{metal11} to any purely poloidal, axisymmetric magnetic field. We illustrate the general theory by calculating explicitly the ellipticity of a star with mixed dipole, quadrupole, and octupole poloidal fields. Lastly, in Sec. 5, we summarize our results and discuss how to constrain the relative weighting of multipoles from current gravitational wave upper limits and future gravitational wave detections.

\section{General formalism}

In the absence of a magnetic field, the star is spherically symmetric and in hydrostatic equilibrium. Let $(r,\theta,\phi)$ be spherical polar coordinates, with $r$ expressed in units of the stellar radius $\rstar$, so that it is dimensionless. To make contact with previous work \citep{metal11,mm12}, we adopt the idealised density profile

\be \rho =\rho_c (1-r^2),\ee
where $\rho_c=15\mstar/(8\pi\rstar^3)$ is the density at the centre, and $\mstar$ is the stellar mass. We emphasize that this is a particular, simple choice of density profile, chosen to render the following calculations tractable, rather than motivated directly by observations or the theory of stellar structure, but it does approximate the $n=1$ polytrope reasonably well \citep{metal11}.

Any axisymmetric magnetic field can be written as \citep{c56}

\be {\bf{B}}=B_0[\eta_p\nabla\alpha(r,\theta)\times\nabla\phi + \eta_t\beta(r,\theta)\nabla\phi],\ee
where $B_0$ is the surface field strength at the equator, and $\eta_p$ and $\eta_t$ are dimensionless parameters defining the relative strengths of the poloidal and toroidal components respectively. The stream function $\alpha(r,\theta)$ can always be factorised into radial and polar parts, i.e., $\alpha(r,\theta)=f(r)\Theta(\theta)$. In addition, the scalar function $\beta$ must be a function of $\alpha$, so that the magnetic force [$\propto(\nabla\times{\bf{B}})\times{\bf{B}}]$ does not have an azimuthal component, which cannot be balanced in hydromagnetic equilibrium.

The magnetic energy density is $\lesssim 10^{-6}$ of the gravitational energy density and hence the pressure $p$. Even in a magnetar, the magnetic force on the star can be treated as a perturbation on a background hydrostatic equilibrium. Thus, we write the hydromagnetic force balance equation as\footnote{Throughout this paper, we use SI units: 1 T $= 10^{4}$ G.}

\be \frac{1}{\mu_0}(\nabla\times {\bf{B}})\times {\bf{B}} = \nabla \delta p +\delta \rho\nabla\Phi,\ee
to first order in $B^2/(\mu_0 p)$ and in the Cowling approximation ($\delta\Phi=0$, where $\Phi$ is the gravitational potential). Note that we do \emph{not} require the density perturbation $\delta\rho$ to be a function purely of the pressure perturbation $\delta p$ (the barotropic assumption). Therefore, we do not restrict the relative strengths of the poloidal and toroidal components of the magnetic field. We only require the following properties:

\begin{enumerate}[labelindent=\parindent,leftmargin=*]
\item the field is cylindrically symmetric about the $z$-axis;
\item the poloidal component is continuous with a purely poloidal field outside the star (so there are no surface currents);
\item the toroidal component is confined to some region inside the star (since the external field has no toroidal component);
\item the current density remains finite and continuous everywhere in the star and vanishes at the surface (since we assume the external field to exist in vacuo, neglecting magnetospheric currents).
\end{enumerate}
All these requirements can be satisfied by choosing a suitable $\alpha$.

This approach differs from that taken by previous authors \citep{hetal08,lj09,cfg10}, who pre-specified a barotropic equilibrium model and then solved for the magnetic field configuration. Because our star is non-barotropic, i.e., because density (background plus perturbation) is not purely a function of pressure, our magnetic field is not constrained by the stellar equation of state, and the toroidal and poloidal components are separately adjustable [they do not need to obey any relations to ensure $\delta\rho=\delta\rho(\delta p)$]. In other words, we stipulate the form of the magnetic field we wish to investigate (eventually to be determined from observational constraints) and solve for the neutron star structure (strictly speaking, the part controlling the ellipticity) that accommodates it. No particular physical stratification mechanism is specified; it is assumed to be whatever is needed to accommodate the chosen field.

We characterize the magnetic deformation of the star by its ellipticity $\epsilon$, defined as

\be \epsilon=\frac{I_{zz}-I_{xx}}{I_0},\ee
where $I_0$ is the moment of inertia of the spherical star, and the moment-of-inertia tensor is given by

\be I_{jk}=\rstar^5\phantom{+}\int d^3{\bf{x}} [\rho(r) + \delta\rho(r,\theta)](r^2\delta_{jk}-x_jx_k),\ee
with the integral covering the interior, $r\leqslant 1$.

We calculate $\delta\rho$ by taking the curl of both sides of Eq. (3). Matching the $\phi$ components, we find

\be \frac{\partial\delta\rho}{\partial\theta} = -\frac{r}{\mu_0\rstar}\frac{dr}{d\Phi}\{\nabla\times[(\nabla\times{\bf{B}})\times{\bf{B}}]\}_{\phi}.\ee
Given ${\bf{B}}$, Eq. (6) can be integrated up to an arbitrary function of $r$ (which does not contribute to the ellipticity). Equations (4)--(5) are subsequently evaluated to give $\epsilon$.

\section{Poloidal-toroidal fields}

In this section, we briefly review the results of \citet{metal11} for a dipole-plus-toroidal field (Sec. 3.1). Then we apply the same method to calculate $\epsilon$ for a poloidal-toroidal field with a quadrupole poloidal component (Sec. 3.2).

\subsection{Dipole plus toroidal field}

\citet{metal11} considered dipole poloidal-plus-toroidal magnetic field configurations. Such fields are broadly representative of the output of numerical simulations \citep{bn06,bs06,b09}. The poloidal flux function is taken to be $\alpha_1=f_1(r)\sin^2\theta$, so that continuity of the poloidal component with an external dipole field is ensured.\footnote{Anticipating Sec. 4, where we discuss arbitrary multipoles, we label all variables with a subscript denoting multipole number $l$.} The function $f_1(r)$ is arbitrary, in principle. For simplicity, \citet{metal11} assumed a polynomial in $r$. One possible choice is $f_1(r)=(35/8)[r^2-(6/5)r^4+(3/7)r^6]$, ensuring that all the continuity and regularity conditions for the field and current in Sec. 2 are fulfilled.\footnote{We discuss briefly the possibility of a four-term polynomial for $f(r)$ in Appendix A.} The toroidal flux function is chosen to be $\beta_1(\alpha_1)=(\alpha_1-1)^2$ for $\alpha_1\geqslant 1$ and $\beta_1(\alpha_1)=0$ elsewhere, so that the toroidal field is confined to the region around the neutral line, where $\alpha_1$ exceeds unity.

Following the procedure in Sec. 2, the ellipticity is calculated to be \citep{metal11,mm12}

\be \epsilon_1 = 5.63\times 10^{-6} \left(\frac{B_{\mathrm{max}}}{10^{11}\textrm{T}}\right)^2 \left(\frac{\mstar}{1.4M_\odot}\right)^{-2}\left(\frac{\rstar}{10^4\textrm{m}}\right)^4\left(1-\frac{0.351}{\Lambda}\right),\ee
where $\Lambda$ is the ratio of the poloidal component's energy to the total magnetic energy\footnote{Magnetic energies are implicitly taken to be \emph{internal}; i.e., we integrate ${\bf{B}}^2$ only to the surface of the star.} ($\Lambda=0$ is a purely toroidal field and $\Lambda=1$ is purely poloidal), and $\bmax$ is the maximum surface field strength (i.e., at the poles). Note that we work in SI units, 1 T = $10^4$ G, 1 J = $10^7$ ergs.\footnote{The lead author apologizes for this cherished idiosyncrasy, which Lasky and Melatos do not share as strongly.} In previous papers, Eq. (7) was combined with gravitational wave upper limits to place bounds on $\Lambda$ for the Crab pulsar, the Cassiopeia A central compact object, newly born magnetars in the Virgo cluster \citep{metal11}, and millisecond pulsars \citep{mm12}.

\subsection{Quadrupole-poloidal-plus-equatorial-toroidal field}

We now calculate the ellipticity due to a mixed poloidal-toroidal field, where the poloidal component is a quadrupole and the toroidal component remains the same as in Sec. 3.1 (localised around the equator). Pulsar observations tell us that neutron star magnetic fields are largely dipolar \citep{cm11}, but we assume that the poloidal component is purely quadrupolar as a first step, in order to understand the effects of higher multipoles on $\epsilon$. Outside the star, the field takes the following form:

\be {\bf{B}}_{\mathrm{ext}}=\frac{B_0}{r^4}[(3\cos^2\theta - 1)\er - 2\sin\theta\cos\theta\etht].\ee
In order to express the field in the form given by Eq. (2), the poloidal flux function must take the form of

\be \alpha_2=f_2(r)\sin^2\theta\cos\theta.\ee

Following Sec. 3.1, suppose $f_2(r)$ is a polynomial. As before, we must first ensure that the current

\be \nabla\times{\bf{B}}\propto \left(\frac{f''}{r}-\frac{6f}{r^3}\right)\sin\theta\cos\theta\ee
is well-behaved as $r\rightarrow 0$, requiring the polynomial to be of degree three or higher. Next, we must ensure that the normal and tangential components of the field and the current are continuous at $r=1$. Therefore, we need at least three terms in the polynomial. We find that $f_2(r)=21\left(r^3-\frac{5}{3}r^4+\frac{5}{7}r^5\right)$ satisfies all the conditions, implying

\be
{\bf{B}}_2 =
   \begin{cases}
      21\eta_p B_0\left[\left(r-\frac{5}{3}r^2+\frac{5}{7}r^3\right)(3\cos^2\theta-1)\er - \left(3r-\frac{20}{3}r^2+\frac{25}{7}r^3\right)\sin\theta\cos\theta\etht\right]+\frac{\eta_t B_0 \beta_2(\alpha_2)\ephi}{r\sin\theta}&\textrm{for }r< 1\\
      %&\\
      \eta_p B_0 r^{-4}[(3\cos^2\theta-1)\er+2\sin\theta\cos\theta\etht]&\textrm{for }r\geqslant 1.
   \end{cases}
\ee

Again, following Sec. 3.1, we choose the toroidal flux function to be

\be \beta_2(\alpha_2) =
\begin{cases}
(|\alpha_2| -1)^2&\textrm{for }\alpha_1\geqslant 1\\
0&\textrm{for }\alpha_1 < 1.
\end{cases}
\ee
Equation (12) confines the toroidal component to the region near the equator that would be occupied by the toroidal component of a dipole field (see Sec. 3.1), instead of to the region around the neutral curves of the quadrupole poloidal field itself, located at $\theta=\cos^{-1}(\pm\sqrt{1/3})$. This is because we expect the toroidal magnetic field generated and/or amplified by differential rotation in a newly born neutron star to be strongest near the equator \citep{bn06,retal01a,retal01b}. We do not suggest here that a neutron star's internal field is predominantly a quadrupole plus a toroidal field at the equator. Indeed, simulations typically generate dipole-dominated configurations \citep{bn06,retal01a,retal01b}. Here, we simply wish to investigate the effects of a quadrupole poloidal component together with some well-motivated toroidal component, and we cite the aforementioned simulations to justify placing the toroidal component at the equator. Furthermore, we find that reasonably simple forms for $\beta_2$ with toroidal field components located around the quadrupole's neutral curves lead to unphysical, discontinuous $\delta\rho$, because the magnetic force that induces $\delta\rho$ is not symmetric about $\theta=\cos^{-1}(\pm\sqrt{1/3})$ [at least for simple forms of $\beta_2$, e.g., a polynomial in $\alpha_2$ as given by Eq. (12)], making it impossible for $\delta\rho$ to vanish smoothly at the torus boundary. We defer the derivation of a quadrupole toroidal field which is mathematically consistent to a future paper.

%For now, we simply wish to highlight the effects of a quadrupole poloidal field and an equatorial toroidal field on a neutron star, as well as to compare directly to the previous case of dipole mixed field, therefore we proceed with Eq. (12).

\begin{figure}
\centerline{\epsfxsize=18cm\epsfbox{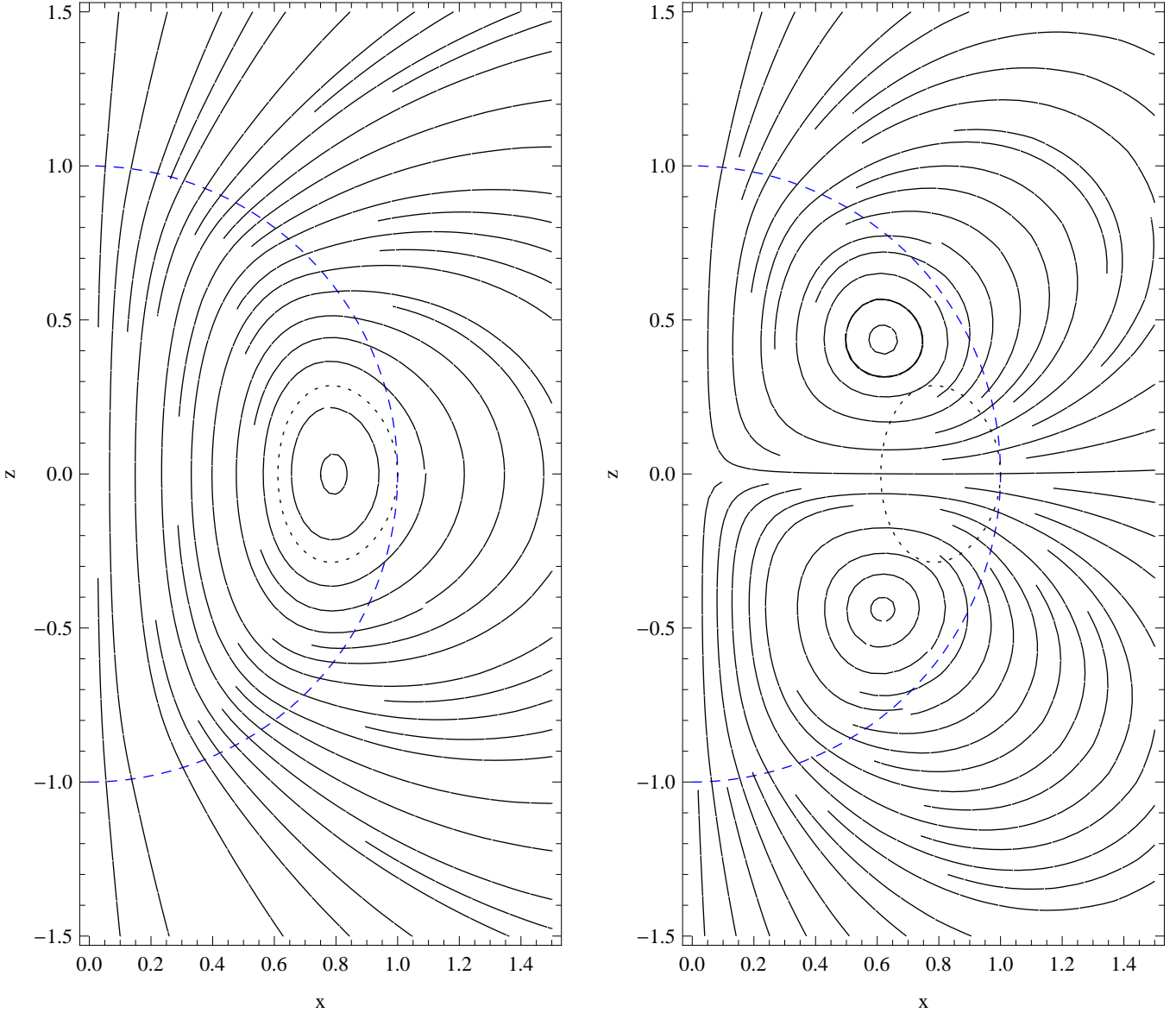}}
 \caption{Magnetic field lines for a dipole (left) and a quadrupole (right). The surface of the star is represented by the blue dashed semicircle. The toroidal field component is confined to the region bounded by the black dotted curve and fills a torus around the $z$-axis. Apparent discontinuities in the field lines are plotting artifacts. The dipole field is north-south antisymmetric, and the quadrupole is north-south symmetric. Also, the dipole only has one neutral curve (where the poloidal field vanishes) at the equator, but the quadrupole has two, located at $\theta=\cos^{-1}(\pm\sqrt{1/3})$.}
 \label{purequadpoltor}
\end{figure}

We sketch the field lines of the dipole in the left-hand panel and those of the quadrupole in the right-hand panel of Fig. \ref{purequadpoltor}. The toroidal field fills the region enclosed by the black dotted curve. In the left-hand panel, the toroidal region is centred around the neutral curve and is bounded by the last dipole field line that fully closes inside the star. The quadrupole (right-hand panel) has one neutral curve in each hemisphere. As stated above, however, since a toroidal magnetic field generated and amplified by differential rotation near the equator is more physically compelling \citep{bn06,retal01a,retal01b}, we choose to confine the toroidal field to the same region as in Sec. 3.1, defined by $|\alpha_1|\leqslant 1$ (instead of the closed quadrupole field lines around the neutral curves).

Upon evaluating Eqs. (4), (5), (11), and (12), we obtain

\be \epsilon_2 = 1.84\times 10^{-6} \left(\frac{\bmax}{10^{11}\textrm{T}}\right)^2 \left(\frac{\mstar}{1.4M_\odot}\right)^{-2}\left(\frac{\rstar}{10^4\textrm{m}}\right)^4\left(1-\frac{0.359}{\Lambda}\right),\ee
where $\bmax$ is the maximum surface field strength (i.e., the surface field strength at the poles). The dependence on $\Lambda$ is similar to the dipole case [Eq. (7)], because the toroidal field is confined to the same region as before. If the toroidal field is localised elsewhere (e.g., around the neutral curves of the quadrupole), the dependence of $\epsilon$ on $\Lambda$ changes. In addition, for given $\bmax$, $\mstar$, and $\rstar$, the quadrupole deforms the star less than the dipole, i.e. $|\epsilon_2|<|\epsilon_1|$, because the magnetic energy of a quadrupole is less than for a dipole with the same $\bmax$.

\section{Composite poloidal fields}

In this section, we discuss neutron star deformation due to a field that is a superposition of multipoles. First, in Sec. 4.1, we derive a general formula for $\epsilon$ due to a multipole of order $l$. Then, in Sec. 4.2, we calculate $\epsilon$ due to a purely poloidal, composite dipole-plus-quadrupole-plus-octupole field as a worked example.

\subsection{General formula for $\epsilon$}

Let us calculate $\epsilon$ for any purely poloidal field that is a superposition of axisymmetric multipoles. We describe the field in terms of spherical harmonics $Y_{lm}(\theta, \phi)$, with ${\bf{B}}=\sum_l \eta_l {\bf{B}}_l$ ($l\geqslant 1$) and

\be {\bf{B}}_l=B_0[g_l(r)Y_{l0}(\theta)\er + h_l(r)Y'_{l0}(\theta)\etht].\ee
Here, $\eta_l$ are constants that determine the weighting of the components, and the prime denotes a derivative with respect to $\theta$.

The radial functions $g_l(r)$ and $h_l(r)$ must be determined for each multipole, such that the conditions in Sec. 2 are fulfilled. Because the spherical harmonics form an orthonormal basis, the total field automatically fulfills the conditions if each multipole fulfills the conditions separately. The task of determining $g_l(r)$ and $h_l(r)$ is made easier by the fact that, according to Eq. (2), $g_l$ and $h_l$ are related through some stream function $\alpha(r,\theta)=f(r)\Theta(\theta)$. By comparing Eq. (14) and the poloidal part of Eq. (2), we find that each multipole obeys $g_l=l(l+1)f/r^2$, $h_l=f'/r$, $\Theta'(\theta)=Y_{l0}(\theta)l(l+1)\sin\theta$, and $\Theta(\theta)=-Y'_{l0}(\theta)\sin\theta$. The field always takes its maximum value, $\bmax=[(l+1)^2(2l+1)/(4\pi)]^{1/2}B_0$, at the poles.

The current density associated with each multipole is

\be \mu_0^{-1}\nabla\times{\bf{B}}=\mu_0^{-1} B_0\left[\frac{f''}{r}-l(l+1)\frac{f}{r^3}\right]Y'_{l0} \ephi.\ee
The polynomial $f(r)$ must contain terms of certain orders if the current is to be well-behaved at the origin; $f(r)$ must also contain at least three terms to fulfill the three conditions at $r=1$ (cf. Appendix A). For the quadrupole, the terms $r^3$, $r^4$, and $r^5$ are sufficient. For the octupole and higher-order multipoles, the terms $r^4$, $r^5$, and $r^6$ are sufficient. We solve for their coefficients from the following boundary conditions at $r=1$: $f''(1)-l(l+1)f(1)=0$; $lf(1)=-1$; and $f'(1)=1$. As noted in Sec. 2, the conditions ensure that the current vanishes at the surface and that the magnetic field is continuous there (i.e., zero surface current).

Upon substituting Eq. (14) into Eq. (6) and integrating, we obtain

\be \frac{-\mu_0\rstar}{B_0^2}\left(\frac{d\Phi}{dr}\right)\delta\rho = \sum_l\sum_{k}\frac{\eta_l\eta_{k}}{r}\left[\frac{d(rh_l)}{dr}-g_l\right]h_{k}Y'_{l0}(\theta)Y'_{k0}(\theta)+\eta_l\eta_{k}\frac{d}{dr}\left\{\left[\frac{d(rh_l)}{dr}-g_l\right]g_{k}\right\}\int^\theta d\theta' Y'_{l0}(\theta')Y_{k0}(\theta'),\ee
where the last integral is an indefinite integral of $Y'_{l0}(\theta')Y_{k0}(\theta')$ over the dummy variable $\theta'$. Strictly speaking, an arbitrary function of $r$ can be added to $\delta\rho$, but it does not alter the mass quadrupole moment, which is the focus of the paper. Substituting $\delta\rho$ from Eq. (16) into Eq. (5), we obtain

\be \epsilon = -\frac{B_0^2 \rstar^4}{4\mu_0 I_0}\sum^{l_\mathrm{max}}_{l=1}\sum^{k_\mathrm{max}}_{k=1} \eta_l \eta_k \sqrt{(2l+1)(2k+1)}(\epsilon_{a,lk}+\epsilon_{b,lk}),\ee
with

\be \epsilon_{a,lk}=\int_{-1}^{1} dx P'_l(x)P_{k}'(x)(1-x^2)(1-3x^2)\int_0^1 dr \left(\frac{dr}{d\Phi}\right)r^3\left[\frac{d(rh_l)}{dr}-g_l\right]h_{k},\ee
\be \epsilon_{b,lk}=\int_{-1}^{1} dx \left[\int^x dy P_l(y)P_{k}'(y)\right](1-3x^2)\int_0^1 dr \left(\frac{dr}{d\Phi}\right)r^4\frac{d}{dr}\left\{\left[\frac{d(rh_l)}{dr}-g_l\right]g_{k}\right\},\ee
where $P_l(x)=(2^l l!)^{-1} d^l(x^2-1)^l/dx^l$ is the Legendre polynomial of order $l$. There are no cross terms between the multipoles in $(\nabla\times {\bf{B}})\times {\bf{B}}$ (i.e., both $\epsilon_{a,lk}$ and $\epsilon_{b,lk}$ vanish when $l\neq k$), except between $l$ and $l\pm 2$. For example, this means that, if the field consists only of a dipole and a quadrupole, each multipole can be treated separately, and the total force and $\epsilon$ are simple sums of the individual contributions.

To understand the correlation between $\epsilon$ and magnetic energy, we first relate the weights $\eta_l$ to $\bmax$ and to the total magnetic energy. Magnetic field energy is defined to be

\be E=\int_{r\leqslant 1} dV \frac{{\bf{B}}^2}{2\mu_0} = \sum_l E_l,\ee
where the energy in the $l$-th multipole is

\be E_l=\frac{2\pi\rstar^3\blmax^2}{(2l+1)(l+1)^2\mu_0}\int_0^1 dr\phantom{i} r^2[g_l^2+l(l+1)h_l^2],\ee
and $\blmax$ is the maximum surface field strength of that multipole. If $g_l$ and $h_l$ are polynomials with three terms in $r$ (to satisfy the three boundary conditions in Sec. 2), it can be shown by explicit calculation that the polynomials with the lowest allowed orders that satisfy the conditions in Sec. 2 (i.e., a polynomial with $r^3$, $r^4$, and $r^5$ terms for the quadrupole and a polynomial with $r^4$, $r^5$, and $r^6$ terms for octupole and higher) maximise $E_l$. We therefore assume throughout the rest of the paper that $g_l$ and $h_l$ are three-term polynomials with the lowest allowed orders.

\begin{figure}
\centerline{\epsfxsize=16cm\epsfbox{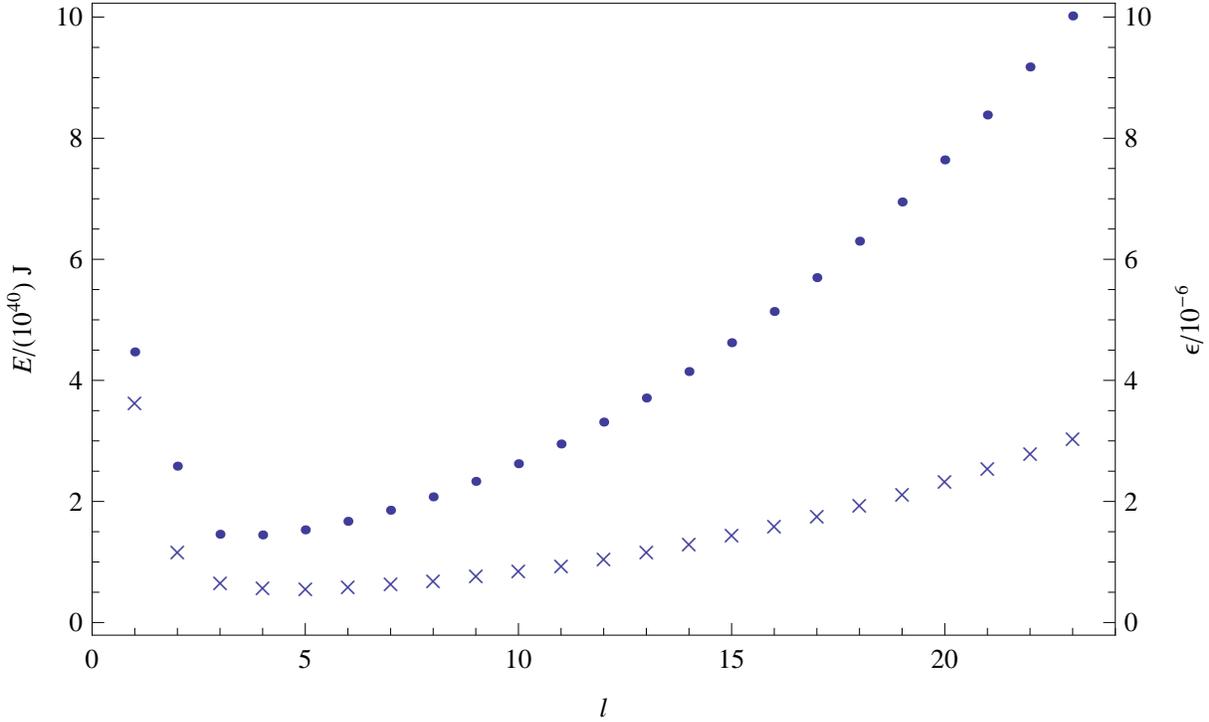}}
 \caption{Magnetic energy $E_l$ (dots, left axis) and ellipticity $\epsilon$ (crosses, right axis) for purely poloidal multipoles, considered separately, as functions of the multipole order $l$. We assume that the star has mass $\mstar=1.4\msun$, radius $10^4$ km, and $\blmax=10^{11}$ T. Ellipticity is directly proportional to $\blmax^2 \rstar^4$ and inversely proportional to $\mstar^2$.}
 \label{energyellvsl}
\end{figure}

In Fig. \ref{energyellvsl}, we plot the magnetic energy and $\epsilon$ for single-multipole fields as functions of multipole order $l$, for a star of mass $1.4\msun$, radius $10^4$ m, and $\blmax=10^{11}$ T. We see that $\epsilon$ is generally proportional to $E$, and that, for a given $\blmax$, stellar mass, and radius, $l=4$ has the lowest energy and $l=5$ has the lowest $\epsilon$; $E$ increases monotonically for $l>4$, and $\epsilon$ increases monotonically for $l>5$.

\subsection{Worked example: dipole plus quadrupole plus octupole}

To illustrate the theory in Sec. 4.1, let us evaluate the specific case ${\bf{B}}=\sum_{l=1}^3 \eta_l {\bf{B}}_l$, i.e., a superposition of purely poloidal dipole, quadrupole, and octupole components. For the dipole, quadrupole, and octupole field energies, we find

\be E_1=\frac{59}{33}\frac{\pi B_{\mathrm{1, max}}^2 \rstar^3}{\mu_0},\ee
\be E_2=\frac{31}{30}\frac{\pi B_{\mathrm{2, max}}^2 \rstar^3}{\mu_0},\ee
\be E_3=\frac{449}{770}\frac{\pi B_{\mathrm{3, max}}^2 \rstar^3}{\mu_0}.\ee

Now let us calculate $\epsilon$ for a linear superposition of $l=1,2,3$ multipoles. From Eq. (14) and Eqs. (18)--(20), after some algebra, we find

\be \epsilon = 3.65\times 10^{-6} \left(\frac{B_{\mathrm{1, max}}}{10^{11} \textrm{T}}\right)^2 \left(\frac{\mstar}{1.4M_\odot}\right)^{-2}\left(\frac{\rstar}{10^4\textrm{m}}\right)^4\left[1+a_1\left(\frac{E_3}{E_1}\right)^{1/2}+\frac{a_2 E_3}{E_1}+\frac{a_3 E_2}{E_1}\right],\ee
with $a_1=2.426$, $a_2=0.567$, and $a_3=0.559$. There are no cross terms between $l=1$ and $l=2$, so the contribution of the quadrupole to the total $\epsilon$ is a simple linear term.

\begin{figure}
\centerline{\epsfxsize=15cm\epsfbox{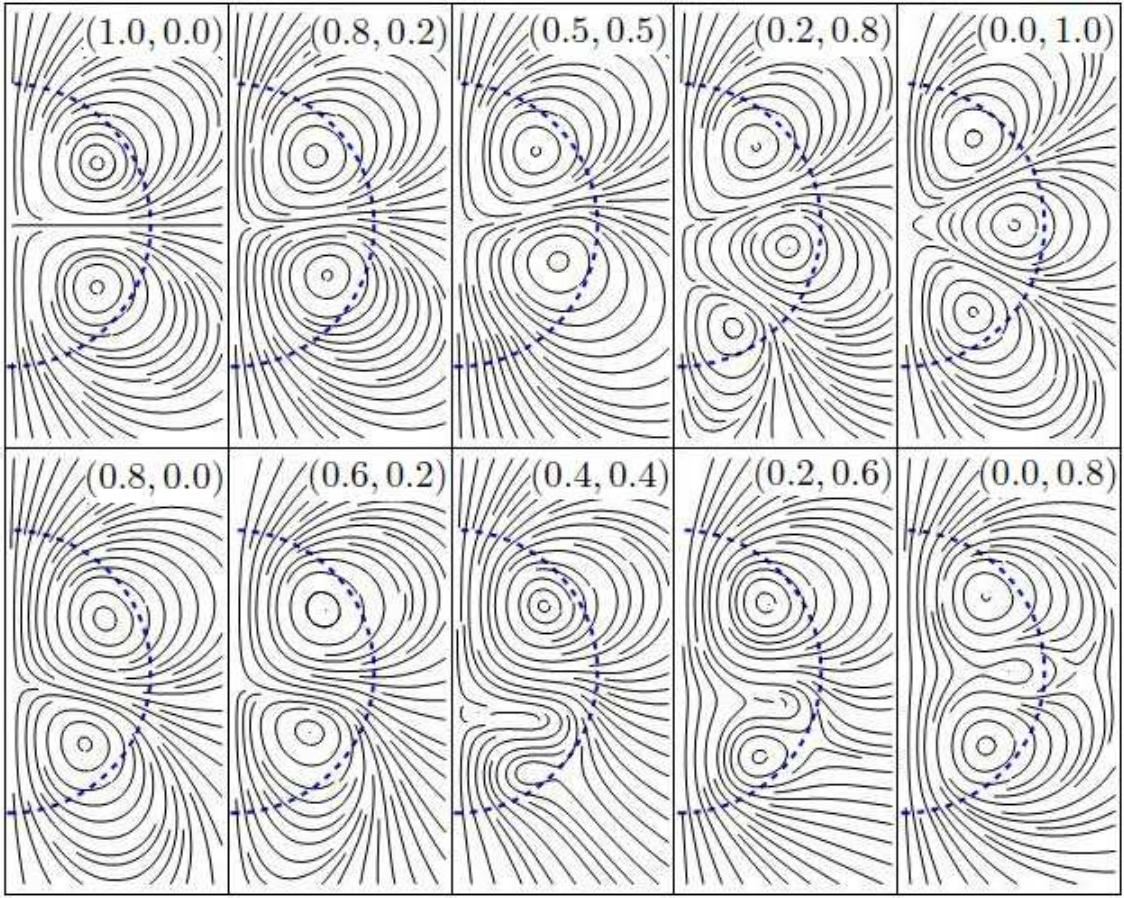}}
 \caption{Magnetic field lines for a superposition of dipole, quadrupole, and octupole fields with different values of $(E_2/E_{\mathrm{tot}},E_3/E_{\mathrm{tot}})$ (label in each panel), where $E_2$, $E_3$, and $E_{\mathrm{tot}}$ are the energies of the quadrupole and octupole components and the total energy respectively. The surface of the star is represented by the blue dashed semicircle. Apparent discontinuities in the field lines are plotting artifacts.}
 \label{dipquadoct}
\end{figure}

For situations where the total magnetic energy $E_{\mathrm{tot}}$ is known (e.g., constrained using SGR giant flare observations), we express $\epsilon$ in terms of $E_2/E_{\mathrm{tot}}$ and $E_3/E_{\mathrm{tot}}$:

\be \epsilon = 4.64\times 10^{-7} \left(\frac{E_{\mathrm{tot}}}{10^{40} \textrm{J}}\right) \left(\frac{\mstar}{1.4M_\odot}\right)^{-2}\left(\frac{\rstar}{10^4\textrm{m}}\right)\left[1+b_1\left(1-\frac{E_2}{E_{\mathrm{tot}}}-\frac{E_3}{E_{\mathrm{tot}}}\right)+b_2\left(1-\frac{E_2}{E_{\mathrm{tot}}}-\frac{E_3}{E_{\mathrm{tot}}}\right)^{1/2}\left(\frac{E_3}{E_{\mathrm{tot}}}\right)^{1/2} +\frac{b_3 E_2}{E_{\mathrm{tot}}}\right],\ee
with $b_1=0.762$, $b_2=4.275$, and $b_3=-1.526\times 10^{-2}$. We plot field lines for some representative combinations of $E_2/E_{\mathrm{tot}}$ and $E_3/E_{\mathrm{tot}}$ in Fig. \ref{dipquadoct}. The pure multipoles are north-south symmetric or antisymmetric, so a superposition of even (odd) multipoles is always symmetric (antisymmetric), for example, $(0.0,1.0)$ and $(0.0,0.8)$ in Fig. \ref{dipquadoct}. However, when odd and even multipoles are mixed, the $\theta$-dependence of the field is no longer (anti)symmetric about the equator, for example, $(0.5,0.5)$ and $(0.2,0.6)$ in Fig. \ref{dipquadoct}.

\begin{figure}
\centerline{\epsfxsize=18.2cm\epsfbox{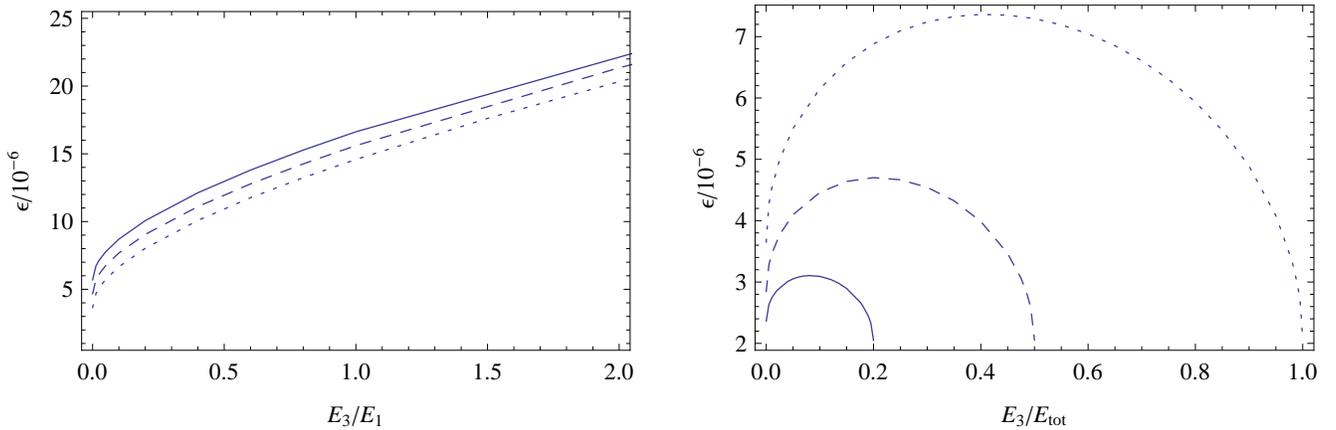}}
 \caption{Ellipticity $\epsilon$ (in units of $10^{-6}$) induced by a magnetic field which is a superposition of dipole, quadrupole, and octupole components, as a function of $E_3/E_1$ (left), the ratio of the energy of the octupole to the dipole, and as a function of $E_3/E_{\mathrm{tot}}$ (right), the ratio of the energy of the octupole to the total energy. Maximum field strength of the dipole $B_{\mathrm{1, max}}$ is kept constant at $10^{11}$ T in the left plot, while total energy is kept constant at $4.47\times 10^{40}$ J (the energy of a pure dipole with $B_{\mathrm{1, max}}=10^{11}$ T) in the right plot. In the left-hand plot, the three curves correspond to different $E_2/E_1$ (the ratio of the energy of the quadrupole to the dipole) equal to 0 (dotted curve), 0.5 (dashed curve), and 1.0 (solid curve). In the right-hand plot, the three curves correspond to different $E_2/E_{\mathrm{tot}}$ (the ratio of the energy of the quadrupole to the total field energy) equal to 0 (dotted curve), 0.5 (dashed curve), and 0.8 (solid curve). Parameters: stellar radius $10^4$ m, stellar mass $1.4\msun$.}
 \label{dqoell}
\end{figure}

In Fig. \ref{dqoell}, we plot $\epsilon$ as a function of $E_3/E_1$ for selected values of $E_2/E_1$ (left-hand panel, for $B_{\mathrm{1, max}}=10^{11}$ T) and as a function of $E_3/E_{\mathrm{tot}}$ for selected values of $E_2/E_{\mathrm{tot}}$ (right-hand panel, for $E_{\mathrm{tot}}=4.47\times10^{40}$ J, the energy of a pure dipole poloidal field with $B_{\mathrm{1, max}}=10^{11}$ T), both for $\mstar=1.4M_\odot$ and $\rstar=10^4$ m. In the left panel, a dipole (i.e., $E_2=E_3=0$) deforms the star into an oblate shape. Adding the quadrupole and octupole induces greater ellipticity, and $\epsilon$ increases monotonically with increasing $E_2/E_1$ and $E_3/E_1$ (left-hand panel). If $E_{\mathrm{tot}}$ is kept constant instead (right-hand panel), we see that $\epsilon$ decreases as $E_2/E_{\mathrm{tot}}$ increases and, for a given $E_2/E_{\mathrm{tot}}$, $\epsilon$ has a maximum when $E_3/E_{\mathrm{tot}}\approx 0.4(1-E_2/E_\mathrm{tot})$. The right-hand panel of Fig. \ref{dqoell} shows that increasing $E_2$ at the expense of $E_1$ reduces $\epsilon$; in other words, the dipole component (which has the highest energy for a given $\bmax$) contributes the most to the magnetic deformation, consistent with Fig. \ref{energyellvsl}.

%\begin{figure}
%\centerline{\epsfxsize=15cm\epsfbox{Annotated3a.eps}}
% %\includegraphics{quadgrid.eps}
% \caption{Magnetic field lines for a superposition of dipole, quadrupole, and octupole fields with different values of $(E_2/E_1,E_3/E_1)$ (label in each panel), where $E_1$, $E_2$, and $E_3$ are the energies of the dipole, quadrupole, and octupole components respectively. The surface of the star is represented by the blue dashed semicircle. Apparent discontinuities in the field lines are plotting artifacts.}
% \label{dipquadoct}
%\end{figure}

%\be E_1=\frac{236}{33}\frac{\pi\eta_1^2 B_0^2 \rstar^3}{\mu_0},\ee
%\be E_2=\frac{62}{15}\frac{\pi\eta_2^2 B_0^2 \rstar^3}{\mu_0},\ee
%\be E_3=\frac{1796}{55}\frac{\pi\eta_3^2 B_0^2 \rstar^3}{\mu_0}.\ee

%Following the procedure outlined in Sections 2 and 3, we obtain the following expression for the octupole field:
%
%\be
%{\bf{B}}_\mathrm{oct} =
%   \begin{cases}
%      &36 B_0\left[\left(r^2-\frac{7}{4}r^3+\frac{7}{9}r^4\right)(12-20\cos^2\theta)\cos\theta\er - 4\left(r^2-\frac{35}{16}r^3+\frac{7}{6}r^4\right)(1-5\cos^2\theta)\sin\theta\etht\right], r< 1\\
%      &\\
%      &\frac{B_0}{r^5}[4(3-5\cos^2\theta)\cos\theta\er+3(1-5\cos^2\theta)\sin\theta\etht),r\geqslant 1.
%   \end{cases}
%\ee
%We reiterate that this expression for the octupole field is not unique; it is only one possible, simple expression which ensures that the field and current are well-behaved in the origin and continuous at the stellar surface.

\section{Discussion}

In this paper, we calculate how the magnetic deformation of a neutron star depends on the orders and relative weightings of the internal magnetic multipoles for a given total magnetic energy. We extend previous calculations \citep{metal11,mm12} to the special case of a quadrupole-poloidal-plus-dipole-toroidal magnetic field, relevant to the poloidal-toroidal twisted torus found by \citet{cetal09}, and to the general case of an arbitrary linear superposition of purely poloidal, axisymmetric multipoles of any order. Our main results are Eqs. (17)--(19), (25), and (26), relating $\epsilon$ to magnetic field strength, stellar mass and radius, and the relative weightings of the multipoles. We show that, in general, $\epsilon$ is proportional to energy $E_\mathrm{tot}$. For single multipoles, $\epsilon$ decreases with $l$ to a minimum at $l=5$, then increases monotonically (Fig. \ref{energyellvsl}). We derive a general formula for $\epsilon$ caused by a superposition of purely poloidal multipoles [Eq. (17)--(19)]. As an explicit example, we calculate $\epsilon$ for a dipole-quadrupole-octupole field as a function of $E_1$, $E_2$, and $E_3$ [Eq. (25)] and as a function of $E_2/E_{\mathrm{tot}}$ and $E_3/E_{\mathrm{tot}}$ [Eq. (26)]. As we see in Eqs. (25)--(26) and in Fig. \ref{dqoell}, a purely poloidal field of any multipole order(s) always deforms the star into an oblate shape and $\epsilon$ increases with $E_2/E_1$ and $E_3/E_1$ [Fig. \ref{dqoell} (left)]. On the other hand, for a fixed $E_{\mathrm{tot}}$, $\epsilon$ decreases as $E_2/E_{\mathrm{tot}}$ increases and, for a particular $E_2/E_{\mathrm{tot}}$, $\epsilon$ reaches a maximum when $E_3/E_{\mathrm{tot}}\approx 0.4(1-E_2/E_\mathrm{tot})$.

Comparing Eq. (13) to Eq. (7), we see that $\epsilon$ is smaller for the quadrupole-poloidal-plus-dipole-toroidal field than for the dipole-poloidal-plus-dipole-toroidal field. Eqs. (7) and (13) indicate that a star with purely poloidal magnetic field is always oblate ($\epsilon>0$). There is, therefore, an upper limit to the ellipticity caused by a purely poloidal field, obtained by setting $\Lambda=1$ in Eqs. (7) and (13), but there is no upper limit on the ellipticity of a prolate star ($\epsilon<0$). In reality, however, there are other limits on $\Lambda$ (and, hence, an upper limit on $\epsilon$ for prolate stars) from stability arguments. \citet{aetal13} demonstrated analytically the stability of a predominantly toroidal field configuration in a non-barotropic neutron star; for a typical magnetar, with $B_{\mathrm{1, max}}=10^{11}$ T, \citet{aetal13} found that $\Lambda\gtrsim 10^{-2}$ is required for stability. Ultimately, however, one needs to conduct time-dependent magnetohydrodynamic simulations to draw convincing conclusions regarding field stability \citep{bn06,letal11,cetal11,lj12}.

%Recent works into the long-term stability of neutron star fields are not yet conclusive. According to \citet{lj12}, a stable neutron star field is likely to be dominated by the poloidal component, in a very narrow range of $\Lambda$. On the other hand, according to \citet{aetal13}, the stable field of a barotropic neutron star is more likely to be dominated by a strong toroidal field, $\sim$ two orders of magnitude stronger than the poloidal field.

X-ray spectra of some magnetars indicate that the surface magnetic field strengths of these magnetars may be greater than the inferred dipole field \citep{gog08,ggo11}. SGR 0418+5729 has an inferred dipole field strength of $\lesssim 7.5\times 10^8$ T \citep{retal10}, but an analysis of the X-ray spectrum, using the Surface Thermal Emission and Magnetospheric Scattering (STEMS) model \citep{getal07}, concluded that a surface field strength of $10^{10}$ T fits the data best [see also \citet{tx12} for an alternative explanation]. \citet{ggo11} postulated higher-order multipole(s) at the surface, which fall away with altitude faster than the dipole, to account for the discrepancy. Substituting $B_\mathrm{1, max}=7.5\times 10^8$ T into Eq. (7), we find $\epsilon=2.05\times 10^{-10}$ for a star with a purely poloidal dipolar magnetic field structure. Adding a quadrupole component with $B_\mathrm{2, max}=10^{10}$ T increases $\epsilon$ to $1.20\times 10^{-8}$; adding an octupole component with $B_\mathrm{3, max}=10^{10}$ T raises $\epsilon$ to $1.08\times 10^{-8}$. While these values of $\epsilon$ are still too small to generate gravitational waves detectable by current-generation interferometers, the presence of superconducting protons \citep{mm12,l13} or quarks \citep{gjs12} in the neutron star core can increase $\epsilon$ by a factor $\sim H_{c1}/\langle B\rangle$, where $H_{c1}$ is the first superconductivity critical field \citep{gas11} and $\langle B\rangle$ is the volume-averaged internal field strength. In SGR 0418+5729, a superconducting interior can raise $\epsilon$ by a factor of 10, increasing the possibility of detection. Hence, any detection of gravitational waves from SGR 0418+5729 will allow us to constrain directly the internal magnetic and material properties of this object. Furthermore, if the orientation of the principal axes of inertia, to be inferred from gravitational wave data, does not match the magnetic inclination angle (the angle between the magnetic and rotation axes), it can be adduced as a compelling evidence for high-order multipoles; as seen in Fig. \ref{dipquadoct}, a superposition of odd and even multipoles is symmetric about the magnetic axis, but not symmetric about the equator.

It is also possible that higher multipoles contribute to the spin down of a newborn magnetar \citep{tcq04,metal11b,b12}. To generate their strong magnetic fields through a dynamo process, it is hypothesized that magnetars are born rotating fast, with period $P\lesssim 1$ ms \citep{td93,tcq04}. \citet{tcq04}, \citet{metal11b}, and \citet{b12} showed how such an object can spin down to $P\sim 1$ s within $\sim 10^2$ s. A pure multipole of order $l$ leads to a spin-down law of the form $\dot\Omega\propto \Omega ^{2l+1}$, where $\Omega$ is the angular velocity of the star, and one has $\dot\Omega\approx \dot\Omega_\mathrm{dipole} (\rstar\Omega/c)^{2l-2}$, where $\Omega_\mathrm{dipole}$ is the spin-down rate of a pure dipole. For mixed multipoles, the dominant multipole is the one with the strongest $|{\bf{B}}|$ at the light cylinder $r=c/\Omega$, and the spin-down law is some intermediate exponent. For submillisecond newborn magnetars, $c/\Omega \sim \rstar$, so the contribution of higher multipoles is non-negligible. The gravitational wave energy emitted during this period also makes the newborn magnetar an excellent candidate for a gravitational wave source \citep{p01,setal05,setal09,metal11}. Note that the gravitational waves emitted during this submillisecond phase contribute to the spin-down law as $\dot\Omega\propto \Omega^5$ \citep{lcc01,c02}. This has important implications for gravitational wave searches, since the phase model (which must be known accurately to perform phase-coherent integrations) depends on braking law.

%\citet{tcq04} did not calculate the spin down due to a quadrupole or octupole, but their results suggest that, as multipole order $l$ increases, the spin-down time-scale also increases, i.e., higher multipole fields on their own exert less torque. This does not mean that higher multipoles do not contribute to the newborn magnetar's spin down. Calculating the effects of higher multipoles on a newborn magnetar's spin down is an interesting and very relevant problem, since it may rule out dynamo models of neutron star field generation/amplification. We defer the self-consistent calculation of these effects to a future paper.

\citet{pbh12a,pbh12b} conducted two-dimensional simulations of a rotating, slowly twisting magnetar magnetosphere. They obtained a series of explosive reconnection events and increases in spin-down torque, which can explain the features observed in the 1998 August 27 and the 2004 December 27 giant flares of SGR 1900+14 and SGR 1806$-$20. Their simulation started from an axisymmetric, north-south-symmetric dipole. As seen in Fig. \ref{dipquadoct}, the addition of quadrupolar and/or octupolar components breaks the hemispherical symmetry, raising the probability of reconnection by complicating the field and simultaneously adding to the reservoir of available magnetic energy. As future work, it will be interesting to ask if a composite $|l|\leqslant 3$ field changes the conclusions of \citet{pbh12a,pbh12b}. If so, then magnetar bursts and giant flares offer another independent way to constrain a magnetar's magnetic geometry from external observations.

%In fact, magnetars in quiescence (i.e., when not bursting or flaring) are generally not expected to be good sources of gravitational waves, because of their long rotational periods.

%Next, comparing Eq. (31) to Eq. (9), we see also that the deformation caused by a purely poloidal field, even one with quadrupole and octupole components, is less than that caused by the presence of a toroidal component. As seen in Eq. (11), the toroidal component, set by the parameter $\Lambda$, deforms the star into an oblate shape rather easily: $\Lambda=0.1$ is enough to give $\epsilon\sim 10^{-5}$ for typical magnetar values. On the other hand, a purely poloidal magnetic field needs a quadrupole or octupole component which is about as strong or stronger than its dipole component before $\epsilon$ changes appreciably. Spin down observations, however, indicate that a neutron star's field is largely dipolar.

\section*{Acknowledgments}

We thank Arthur Suvorov for discussions. We thank the anonymous reviewer for the constructive comments. Alpha Mastrano thanks Kostas Kokkotas for the hospitality shown during his stay at the University of T\"{u}bingen, where this work was completed. This work is supported by an Australian Research Council Discovery Project Grant (DP110103347) and a University of Melbourne Early Career Researcher Grant.

\appendix

\section{Radial factor of the stream function}

Throughout the paper, we postulate that $f(r)$, the radial factor of the magnetic field's separable stream function $\alpha(r,\theta)=f(r)\Theta(\theta)$, takes the form of a polynomial in $r$ with three terms, the minimum needed to fulfill the boundary conditions (ii)--(iv) in Sec. 2.

There is, in fact, no absolute requirement that $f(r)$ only consists of three terms (nor indeed that it be a polynomial). One can add one or more terms whose coefficients are not fixed by the boundary conditions. This allows one to fine-tune the field to match observations, while keeping the polynomial form of $f(r)$ to simplify calculations.

As an example, let us examine the case of a purely poloidal dipole field. In Sec. 3.1, we used $f(r)=(35/8)[r^2-(6/5)r^4+(3/7)r^6]$ \citep{metal11,aetal13}. If we now assume that $f(r)$ takes the form of $f(r)=ar^2+br^4+cr^6+dr^5$, we can reevaluate $\epsilon$ to find its dependence on $d$. We leave $d$ as a free parameter, then solve for $a$, $b$, and $c$ to match the boundary conditions. We find

\be \epsilon_d = 3.754\times 10^{-10} \left(\frac{B_{\mathrm{max}}}{10^{11}\textrm{T}}\right)^2 \left(\frac{\mstar}{1.4M_\odot}\right)^{-2}\left(\frac{\rstar}{10^4\textrm{m}}\right)^4(d^2+47.47d+9.728\times 10^3),\ee
and the field energy $E$ is

\be E=3.247\times 10^{-4}(d^2+81.67+5.507\times 10^3)\frac{\pi B_{\mathrm{1, max}}^2 \rstar^3}{\mu_0}.\ee
The energy $E$ has a minimum at $d\approx -41$ and $\epsilon_d$ has a minimum at $d\approx -24$.

\begin{figure}
\centerline{\epsfxsize=14cm\epsfbox{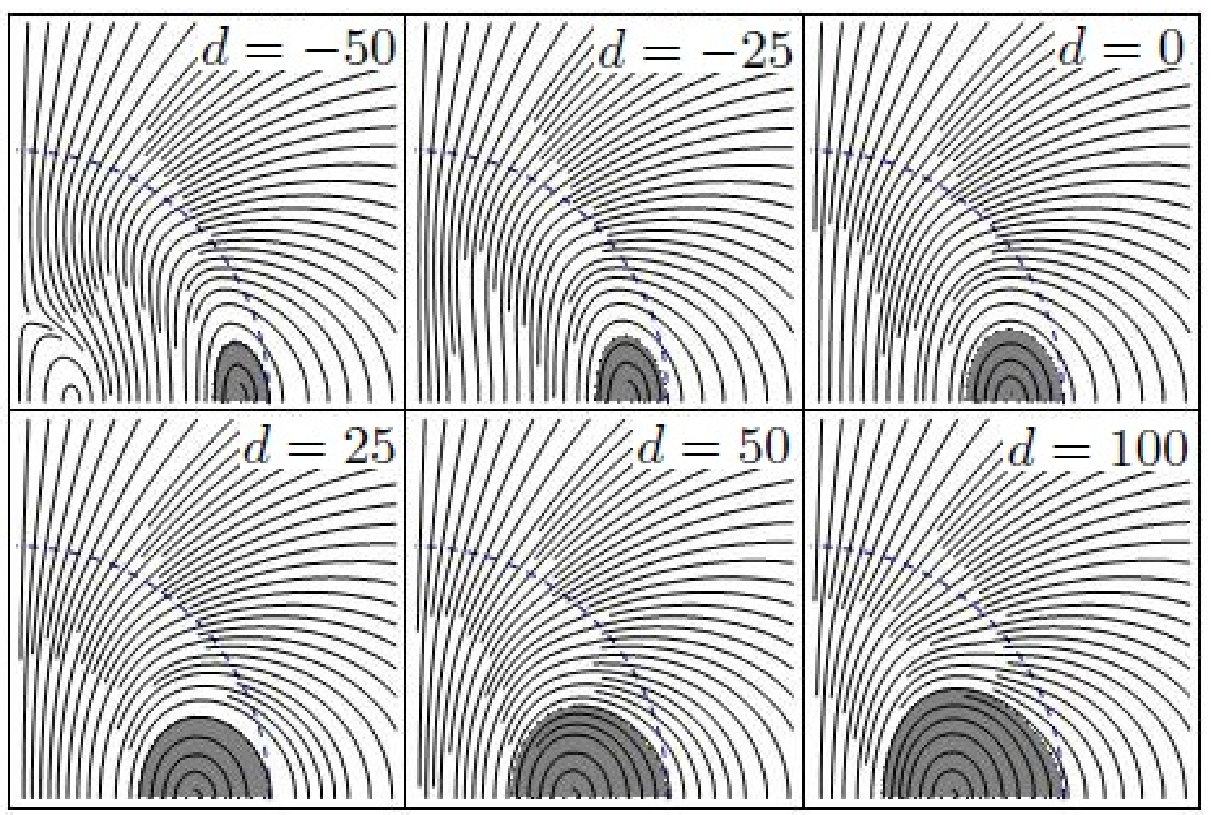}}
 \caption{Field line plots of the dipole poloidal-plus-toroidal fields, where the radial part of the stream function is $f(r)=ar^2+br^4+cr^6+dr^5$, for six different values of the fourth polynomial coefficient $d$. The dashed semicircle represents the stellar surface. The toroidal component, if present, is confined to the area around the neutral curves, represented by the shaded region. Apparent discontinuities in the field lines are plotting artifacts.}
 \label{dip5p}
\end{figure}

We plot the field lines for six different values of $d$ in Fig. \ref{dip5p} [including $d=0$ for comparison]. As Fig. \ref{dip5p} shows, the region occupied by the toroidal field (if it is present) expands as $d$ increases. This raises the possibility of a toroidal field with a relatively weak $|{\bf{B}}|$ possessing a disproportionately large energy, thereby affecting stability and inferences concerning the SGR burst/giant flare energy reservoir. Fig. \ref{dip5p} shows as well that one can have another neutral curve closer to the origin when $d$ is large and negative (e.g., $d=-50$). In theory, a second toroidal field region can exist around this inner neutral curve. These interesting possibilities and calculations will be investigated in a future work.

%\caption{Plot of $f(r)=ar^2+br^4+cr^6+dr^5$ versus $r$ for the dipole, for four different values of the coefficient $d$: $-50$ (dotted curve), $-25$ (dashed-dotted curve), 25 (dashed curve), and 50 (solid thin curve). The thick solid curve represents the case of $f(r)=ar^2+br^4+cr^6$ presented in Sec. 3.1 of this paper and in \citet{metal11}. If the toroidal component is present, it will be localised in the area bounded by $f(r)$ and $f=1$.}

\bsp \label{lastpage}

\end{document}